\documentclass[pof,showpacs,reprintnumbers,amsmath,amssymb,txfonts,pxfonts]{revtex4-1}
\usepackage[utf8]{inputenc}
\usepackage{amsmath}
\usepackage{graphicx} 
\DeclareGraphicsExtensions{.pdf,.png,.jpg,.eps,.ppm}
\usepackage[font=small,width=0.75\textwidth]{caption}
\usepackage{caption}
\usepackage{subcaption}
\usepackage[margin=1in]{geometry}
\usepackage{setspace}
\usepackage[pdftex,bookmarks,bookmarksopen,bookmarksdepth=2]{hyperref}
\hypersetup{colorlinks=true,citecolor=blue}

\begin{document}
\title{A numerical approach for the direct computation of flows including fluid-solid interaction: modeling contact angle, 
        film rupture, and dewetting}

\author{K. Mahady}
\author{S. Afkhami}
\author{L. Kondic}
\affiliation{Department of Mathematical Sciences,
New Jersey Institute of Technology,
Newark, NJ 07102 USA}

\date{\today}
\begin{abstract}

In this paper, we present a computationally efficient method for including fluid-solid interactions
into direct numerical simulations of the Navier--Stokes  equations. This method is found to be as powerful 
as our earlier formulation [J. Comp. Phys., vol. 249: 243 (2015)], while outperforming the earlier method 
in terms of  computational efficiency. The performance and efficacy of the presented method are demonstrated 
by computing contact angles of droplets at equilibrium.  Furthermore, we study the instability of films due to 
destabilizing fluid-solid interactions, and discuss the influence of contact angle and inertial effects on 
film breakup.  In particular, direct simulation results show 
an increase in the final characteristic length scales when compared to the predictions of a linear stability 
analysis, suggesting significant influence of nonlinear effects.
Our results also show that emerging length scales differ,
depending on a number of physical dimensions considered.  

\end{abstract}

\maketitle

\section{Introduction}\label{sec:introduction}

Stability of thin films, in particular at the nanoscale, is
relevant in a variety of applications where film breakup and dewetting are
important.  One example of such applications involves
harnessing instabilities of nanoscale liquid metal films in
order to create arrays of nanoparticles via self- or directed-assembly~\cite{kd_pre09, fowlkes_nano11,
Fowlkes2013, gonzalez2013}.  These methods involve the deposition of thin,
solid metal films, that are subsequently exposed to laser pulses; breakup
leads to the formation of drops, which then solidify to form nanoparticles.  The
applications of such nanoparticles are wide, ranging from solar cells to liquid
crystal displays, among
others~\cite{fleming_phystoday08,maier_natmat03,sun_science00,maier_2007,bader_rmp06,min_natmat08}.
Such nanofilm breakup presents numerous computational challenges that may
result from complex initial geometries~\cite{Roberts2013, Mahadysqw2015};
however, rupture of a simple flat film due to fluid-solid interactions also
demonstrates rich behavior that has been studied extensively in the past (see,
for example,~\cite{sharmaJam1993, sharma_jcis99, sharma_epje03},
or~\cite{cm_rmp09} for a review).  

To model the film breakup and the consequent dewetting phenomena, it is necessary
to include a destabilizing mechanism: if such a mechanism is not included, a continuous
film does not break up.  In this context, the long-wave formulation (L-W)
is usually considered, since it simplifies the underlying mathematical 
model, reduces dimensionality of the problem, and comes at significantly reduced
computational cost.   Liquid-solid interaction forces are often included in the
L-W as  conjoining-disjoining pressure, leading to a prewetted (often called
`precursor') layer in nominally `dry' regions.  This approach effectively
removes the `true' contact line, consequently avoiding the 
non-integrable shear-stress singularity at the moving contact line
(see e.g.~\cite{Schw98,dk_pof07,eggers2005}).  
Another approach to alleviate the moving contact line singularity 
is to relax the no-slip condition and instead assume the presence
of slip at the liquid-solid interface (see e.g.~\cite{Hocking2,Munch2005,afkhami_jcp09,Shikhmurzaev}). 
Slip at the solid surface for fluid-fluid systems has also been studied 
in the context of molecular dynamics simulations (see e.g.~\cite{KBW88,TR89,Thompson1997,Ren2007}). 
Both slip and disjoining pressure approaches have been
extensively used to model a variety of problems including wetting, dewetting,
and
film breakup, in particular in the context of polymer~\cite{seeman_prl01,becker_nat03}
and metal~\cite{ajaev_pof03,trice_prl08,kd_pre09,fuentes_pre11,khenner_pof11,Fowlkes2013,gonzalez2013}
films.  

The  L-W, however, has its limitations, in particular regarding 
the requirements of small interfacial slopes and negligible inertial effects.   
For example, the L-W may not be sufficient to provide quantitatively
precise  predictions regarding instability development for metal films at 
nanoscales, where contact angles are large, and inertial effects 
considerable~\cite{habenicht_05,Habenicht08,Fuentes11,afkhami-kondic-2013}. 
Regarding the small slope requirement, it is often argued that the results of
the L-W are reasonably accurate even if this requirement is not 
strictly satisfied; in addition, various improvements of the L-W are
available, see, e.g.~\cite{snoeijer06}.  Still, since it is not always possible to compare results of
simulations directly to experiments, it is difficult to judge to which degree 
the L-W based results quantitatively describe the process of thin film breakup.   
Regarding inertial effects, although the L-W can be extended to include
them (see~\cite{oron_rmp97, craster_rmp09} for reviews of this topic), the resulting formulations are not
straightforward, and are limited in the range of Reynolds numbers that can be considered.

To be able to accurately model  the problems where the assumption of small slopes is not 
satisfied, or where inertial effects are important, it is necessary to go beyond
the L-W, and consider direct solutions of Navier--Stokes  equations, that also 
include fluid-solid interaction forces. It is therefore essential to design a computational method to include the 
following: (i) precise resolution of long and short range fluid-solid interactions; 
(ii) allow non-negligible inertial effects and large contact angles; 
(iii) provide spatially and temporally converged solutions with a reasonable computational cost.

In our earlier work~\cite{MahadyvdW15}, we developed a model for inclusion of
fluid-solid interaction forces in a Navier--Stokes  solver.  However, that model
requires significant computational effort, and it is not practical for use in
more complex scenarios, such as three dimensional (3D) film breakup.   In the
current paper,  we present a computational approach that is significantly more
efficient.  The present approach permits the analysis of 
complex evolution of the
rupturing and dewetting process. 
Our results reveal that the film initially evolves in accordance to the predictions of the 
linear stability analysis (LSA) based on the L-W, but 
departs from the LSA prediction beyond the onset of the instability.
We also discuss the influence of the number of physical dimensions considered, by 
showing differences in the rupture process between two and three dimensions (2D and 3D).
Despite the fact that a variety of other computational methods have been considered
in the context of wetting/dewetting (see e.g.~\cite{Jacqmin1999,Jacqmin2000,Yue2010}), 
to our knowledge, the numerical scheme presented in this work provides 
the only available framework that satisfies all three requirements listed above
to quantitatively describe the dynamical evolution of thin film instability including 
rupture.

This paper is organized as follows. We begin in Sec.~\ref{sec:model}
by giving an overview of the computational framework and a brief review of  the L-W. 
We also briefly discuss the LSA of the L-W equations of an initially flat film on a
substrate. In Sec.~\ref{sec:results_vdwreducedpressure}, we compare the results for the computed
contact angle with available results. Then, we present the results
of the film instability in linear as well as nonlinear regimes in 2D.  
We study the underlying physical process, concentrating in particular on the role of the inertia and
contact angle to better understand the regimes of the validity of the LSA.
Finally, we present simulations of nonlinear evolution and breakup of 3D films
and contrast the results with 2D ones. The main conclusions are summarized in Sec.~\ref{sec:conclusions}.

\section{Model}\label{sec:model}

\subsection{Derivation}

We consider a solid phase occupying a half-infinite region
$y<0$, above which there is a region occupied by two fluids that we conventionally
refer to as a vapor phase (variables subscripted $v$) and a liquid phase
(variables subscripted $l$). The words `vapor' and `liquid' are used purely
conventionally, and do not necessarily signify anything about the physical
properties, except that the two phases may differ in terms of their densities
and viscosities.

Each particle of fluid phase $i$ interacts with the solid substrate by
means of a Lennard-Jones type potential~\cite{Israelachvili}
\begin{equation}\label{eq:particlevdwpotential}
  \phi_{is} = K_{is}^\ast \left( \left( \frac{\sigma}{r}\right)^p 
  	-\left( \frac{\sigma}{r}\right)^q \right),
\end{equation}
where $r$ is the distance between the two particles, and $K_{is}^\ast$ is
the scale of the potential, such that Eq.~\eqref{eq:particlevdwpotential}
has a minimum $K_{is}^\ast/4$ at $r=(p/q)^{1/(p-q)}\sigma$. 
We can obtain the total potential energy in phase $i$ due to this interaction
by the following expression
\begin{equation}\label{eq:totalenergy}
  \Phi_{is} = n_i \int_{-\infty}^{\infty} \int_{-\infty}^{0} \int_{-\infty}^{\infty} \phi_{is} n_s dxdydz,
\end{equation}
where $n_i$ and $n_s$ are the densities of particles in fluid phase $i$ and  
the solid substrate, respectively. Performing the 
integration, we obtain
\begin{equation}\label{eq:interaction}
    \Phi_{is}(y) = \mathcal{K}_{is}\left[\left(\frac{h^\ast}{y}\right)^m -
	\left(\frac{h^\ast}{y}\right)^n\right] = \mathcal{K}_{is} F(y),
\end{equation}
where
\begin{equation}\label{eq:strength}
	\mathcal{K}_{is} = 2\pi n_i n_s K^\ast_{is}\sigma^3 \left(\frac{\left[(p-2)(p-3)\right]^{q-3}}{\left[(q-2)(q-3)\right]^{p-3}}\right)^{\frac{1}{p-q}},
\end{equation}
\begin{equation}\label{eq:equifilm}
	h^\ast = \left[\frac{(q-2)(q-3)}{(p-2)(p-3)}\right]^\frac{1}{p-q}\sigma;\quad\quad  m = p-3, \quad n=q-3.
	\end{equation}
Equation \eqref{eq:interaction} gives the total potential per unit volume in fluid
phase $i$ due to the interaction with the solid substrate.
The quantity $h^\ast$ is conventionally referred to as the equilibrium film 
thickness in the literature~\cite{dk_pof09, dk_pof07}, and we will continue
using this convention here. 
The term equilibrium film thickness arises because $h^\ast$ is the thickness of a film
that represents an energetic minimum due to Eq.~\eqref{eq:interaction}, and in this model completely
wets the surface of the substrate.

We consider the inclusion of the potential in Eq.~\eqref{eq:interaction} in the
Navier--Stokes  equations for two phases. 
For clarity, we will refer to two phases  
as the liquid phase (subscript $l$), and the vapor phase (subscript $v$), i.e.~$i=l,v$,
although the present formulation applies to any two fluids. 
In order to do this, we introduce a characteristic function $\chi(x,y,z)$, which takes the value
of $1$ inside of the liquid phase, and $0$ inside the vapor phase. 
The interface between these two phases is assumed to be sharp, so that $\chi$
changes discontinuously at the interface.
The governing equations consequently become
\begin{equation}\label{eq:navierstokes}
  \rho(\chi)\frac{D \textbf{u}}{Dt} = -\nabla p + \nabla\cdot\left[ \mu (\chi)\left(\nabla
    \textbf{u} + \nabla\textbf{u}^\top\right)\right]
    + \gamma\kappa\delta_s\textbf{n} - \mathcal{K}(\chi)\nabla F(y), 
\end{equation}
\begin{equation}\label{eq:incompressibility}
  \nabla\cdot\textbf{u} = 0.
\end{equation}
Here, $\rho$ is the (phase dependent) density, $\mu$ is the viscosity (phase dependent), 
$p$ is the pressure, 
and $\textbf{u}$ is the velocity vector.
Also, $D/Dt = \partial_t  + \textbf{u}\cdot \nabla$ is the material
derivative.
The surface tension is included as a singular body force~\cite{Brackbill92}.
Here $\gamma$ is the coefficient of surface tension, $\kappa$ is the interfacial curvature,
$\delta_s$ is a delta function centered on the interface, and $\textbf{n}$ is
a normal vector for the interface pointing out of the liquid.
The quantities depend on $\chi$ via the following
\begin{align*}
  \rho & = \rho_l\chi + \rho_v(1-\chi),\\
  \mu &= \mu_l\chi + \mu_v(1-\chi),\\
  \mathcal{K} &= \mathcal{K}_{ls}\chi + \mathcal{K}_{vs}(1-\chi).
\end{align*}

In order to nondimensionalize, Eqs.~\eqref{eq:navierstokes}~-~\eqref{eq:incompressibility},
we introduce the length scale $L$, and define the time scale as the capillary time, 
$\tau = \mu_l L/\gamma$, since, 
for the parameter sets considered in this paper the dynamics is dominated by viscous and
surface tension effects.
The dimensionless variables are defined as follows
\begin{equation}
\nonumber
 \tilde{x} = \frac{x}{L};   \quad 	 	\tilde{y} = \frac{y}{L}; \quad  \tilde{z}=\frac{z}{L}; 	\quad		\tilde{t}=\frac{t}\tau;
 \end{equation}
 \begin{equation}
 \nonumber
 \tilde{h}^\ast = \frac{h^\ast}{L}; \quad 	 \tilde{\textbf{u}} = \frac{\textbf{u}\tau}{L}; \quad  \tilde{p} = \frac{L p}{\gamma};\quad   \tilde{\kappa} = L\kappa;\quad
 \tilde{\rho} = \frac{\rho}{\rho_l} ;  \quad \tilde{\mu} = \frac{\mu}{\mu_l};\quad  \tilde{\delta_s} = L\delta_s.
\end{equation}
With these scales, and dropping the tildes, the dimensionless Navier--Stokes  equations become
\begin{equation}\label{eq:NS_nondim}
  (\mbox{Oh}^2)^{-1} \rho \frac{D\textbf{u}}{Dt} = -\nabla p +
    \nabla\cdot\left[
    \mu(\nabla\textbf{u} + \nabla \textbf{u}^T) \right] + \kappa \delta_s
    \textbf{n} - K \nabla F(y),
\end{equation}
where the Ohnesorge number is defined via
$\mbox{Oh}^2  = \mu_l^2/(\rho_l \gamma L)$, with $L$ chosen according to the problem
under consideration, 
and
$$K	
		=  K_{ls} \chi + K_{vs} (1-\chi) 
		= \frac{\mathcal{K}}{\gamma L^3} 
		= \frac{1}{\gamma L^3} 
			\left(\mathcal{K}_{ls} \chi + \mathcal{K}_{vs} (1-\chi)  \right).$$
Equation~\eqref{eq:incompressibility} is invariant with this scaling.
In~\cite{MahadyvdW15}, we solved Eq.~\eqref{eq:NS_nondim} directly.
In that paper, we described two methods that we
here refer to as `body force' methods; in these methods,
we discretized the fluid-solid interaction term differently, but
we found both methods to be substantially similar in their overall properties.
For consistency, when we refer to the `Body Force' (B--F) method in this paper,
we exclusively refer to Method II in \cite{MahadyvdW15}.
As described in our previous work, the fluid-solid term presents a challenge
for body force methods, in that the potential is divergent
as $y \rightarrow 0$.
Consequently, parts of the domain near the substrate require a high mesh resolution 
in order to obtain reasonably accurate results. This requirement significantly limits the use of adaptive
meshes, that are crucial for the purpose of improving the performance
of direct simulations.  In 2D, simple problems are
feasible, however, in 3D, the computational cost of resolving so
much of the domain makes the simulations impractical.

In this section, we show that the computational task is dramatically simplified
by reformulating the body force term in Eq.~\eqref{eq:NS_nondim} as
a force which acts only on the interface.
First, we define 
$$p^\ast = p + K_{ls} \chi F(y) + K_{vs} (1-\chi)F(y),$$
so that
$$
-\nabla p^\ast = -\nabla p - (K_{ls} \chi  + K_{vs}(1-\chi))\nabla F - (K_{vs} - K_{ls})\delta_s \textbf{n} F(y),
$$
where $\delta_s\textbf{n} = -\nabla \chi$, in a distributional sense.
Substituting into Eq.~\eqref{eq:NS_nondim}, we obtain what we refer to as the 
`Reduced Pressure' (R--P) formulation 
\begin{equation}\label{eq:NS_rp}
  (\mbox{Oh}^2)^{-1} \rho \frac{D\textbf{u}}{Dt} = -\nabla p^\ast + \nabla\cdot\left[
    \mu(\nabla\textbf{u} + \nabla \textbf{u}^T) \right] + 
    (\kappa +\varkappa F(y)) \delta_s \textbf{n},
\end{equation}
where $\varkappa=(K_{vs}-K_{ls})$ 
(note that only this difference is relevant, rather than individual values of $K_{vs}$ and $K_{ls}$).

Simple energy arguments~\cite{MahadyvdW15} can be used to show that the
fluid-solid interaction term in Eq.~\eqref{eq:NS_rp} gives rise to an
equilibrium contact angle $\theta_{\text{eq}}$ for the liquid phase given by
\begin{equation}\label{eq:diffKtheta}
  \varkappa = \frac{(1-\cos\theta_{\text{eq}})}{h^\ast}\left(\frac{(m-1)(n-1)}{m-n}\right).
\end{equation}
The exact meaning of $\theta_{\text{eq}}$ requires some clarification.  
For the fluid-solid interaction considered (with conjoining and disjoining components),
there is a film of thickness $h^\ast$ that wets the entire substrate, with a 
smooth transition from the droplet to the equilibrium film.
We measure contact angles by fitting a circular profile
to the droplet profile `away' from the transition region; the angle at which
this circular profile intersects the equilibrium film is taken to be the contact angle.
The angle $\theta_{\text{eq}}$ is the equilibrium contact angle in the following sense:
for a drop at equilibrium with a vanishingly small $h^\ast$,
applying the above fitting procedure yields $\theta_{\text{eq}}$.
For non-vanishing but small $h^\ast$, a drop at equilibrium will have a (slightly) different 
contact angle, which we refer to as $\theta_{\text{num}}$.

In addition to the contact angle, the fluid-solid interaction gives rise
to another phenomenon that we consider in this paper:
the spontaneous rupture of a thin liquid film.
Our account of film rupture comes from~\cite{dk_pof07}, where this 
phenomenon is studied in the context of the L-W. 
For completeness, in Sec.~\ref{L-W}, we outline the L-W based approach for
inclusion of the fluid-solid interaction using the disjoining pressure model.

\subsection{Computational Implementation}
\label{implemnetation}

\begin{figure}[t]
  \centering
  \includegraphics{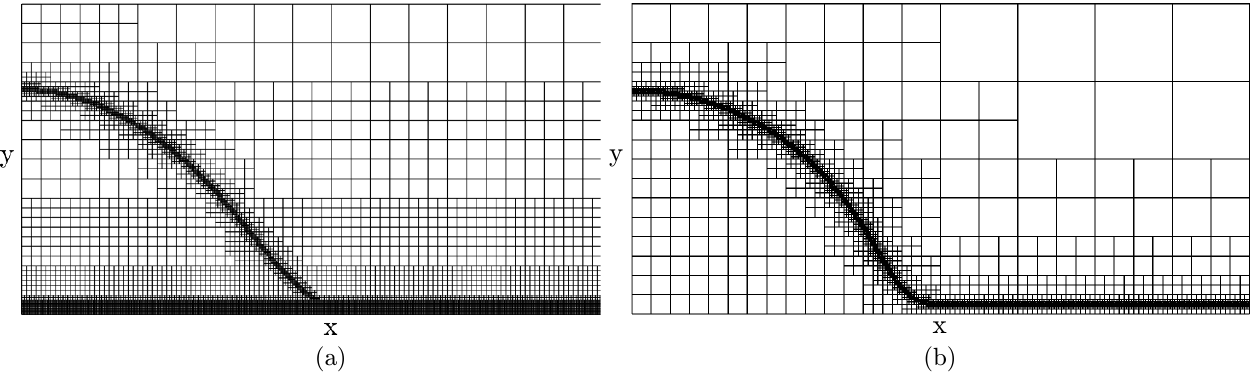}
  \caption{(a) Illustration of a typical adaptive mesh required for the B--F method. 
	       The fluid-solid interaction applies everywhere in the domain,
	       requiring increased resolution along the bottom of 
	       the domain.
	   (b) Illustration of the adaptive mesh used for simulations reported
	   in this paper. 
		The interface is the only portion of the domain that needs to
		be resolved at high
		resolution, due to the fact that the fluid-solid interaction
		is included as an interfacial force.}
  \label{fig:mesh}
\end{figure}

The R--P method discussed in this paper possesses three important strengths.
First, in contrast to conventional methods for implementing the contact angle, 
that impose it essentially as  a constraint 
that the solution needs to satisfy, 
it possesses the most important feature of the B--F method discussed in~\cite{MahadyvdW15}:
it includes long range fluid-solid interaction, and therefore spontaneous film breakup
can occur.   This feature is crucial if one wants to analyze instability of fluid films on nanoscale.  
The second major advantage of the formulation presented in this work is that by
absorbing the entire contribution of the fluid-solid interaction
into a surface force, the main weakness of the B--F method is avoided.
As discussed previously~\cite{MahadyvdW15}, the B--F method significantly
limits the usefulness of adaptive mesh refinement. Since the 
force term in the B--F method applies everywhere in the fluid domain, and is singular as
$y\rightarrow 0$, the entire domain near $y=0$ requires a finer mesh
in order to resolve the force; a typical adaptive mesh for simulations
from~\cite{MahadyvdW15} using the B--F method is shown in
Fig.~\ref{fig:mesh}(a).
In the R--P method, the force due to the fluid-solid interaction is
applied as a surface force; consequently, high resolution is only required
at the interface. 
Figure~\ref{fig:mesh}(b) shows an adaptive mesh
used for simulations using the R--P method in this paper.
A further advantage of the R--P method is that it is relatively
simple to implement, owing to the fact that it can be formulated as a
modification of the curvature in the surface tension term in the Navier--Stokes equations.

We solve Eqs.~\eqref{eq:NS_rp} using the software package 
Gerris~\cite{popinetGerris}, described in detail in~\cite{Popinet2009}.
The mesh consists of a quad tree (in 2D) or an octree (in 3D), that
decomposes the domain into square control volumes, which we refer to as cells (see Fig.~\ref{fig:mesh}). 
We use an adaptive mesh for all simulations, refining the interface to a resolution of $\Delta_{\text{max}}$,
and to lower resolutions far away from the interface.

The interface is tracked using the Volume of Fluid (VoF)
method. The VoF method replaces the characteristic function $\chi$ of Eq.~\eqref{eq:navierstokes}
with its average over each cell. This cell average is referred to as the volume fraction,
$T$, and denotes the fraction of each computational cell occupied by the liquid phase.
The VoF method reconstructs the interface as a sharp interface
which is piecewise linear in each cell. 
Our fluid-solid interaction is included as a modification of the curvature
in the surface tension term described in ~\cite{Popinet2009}.
In this method, the curvature, $\kappa$, is calculated at cell centers, and
our method replaces it with $\kappa + K F(y)$, where 
$y$ is the $y$-coordinate of the midpoint of the interface in an interfacial cell.
The boundary conditions are as follows: on the solid substrate, we impose no-slip and 
$T=1$; on all other boundaries, we impose a homogeneous Neumann boundary condition on all
variables.

\subsection{Long-wave formulation (L-W)}
\label{L-W}

The long-wave formulation (L-W) describes the evolution of the liquid film thickness, $h(x)$
(or $h(x,z)$ in 3D), subject to the well-known assumptions of small interfacial 
slopes (and therefore small contact angles) and negligible inertial effects 
see, e.g.~\cite{oron_rmp97} for an extensive review of the L-W.  
In the present work, we will use the linear stability analysis (LSA) of the L-W  to validate
the simulation results and to highlight the differences between the LSA predictions
and the direct numerical simulations, in particular in the presence of inertia 
and large contact angles. We note that the focus in this paper is not on extensive comparison
between the simulation and the L-W  results. Such a comparison can be found in~\cite{mahady_13}
for spreading and retracting droplets.

In L-W, an additional contribution, 
of the form $\Pi(h) = \mathcal{K}_{is}F(h)$, to the evolution equation 
is made to include the fluid-solid interaction.  Note
that $\Pi(h)$ is of the same form as the fluid-solid interaction term, Eq.~\eqref{eq:interaction}, 
formulated for a flat film. See~\cite{MahadyvdW15} for more details, and~\cite{oron_rmp97,Schwartz-review2002} for reviews;  an extensive discussion regarding inclusion
of disjoining pressure in the L-W equation can be found in~\cite{dk_pof07}.  
With $\Pi(h)$ included, the dimensionless L-W equation is given by 
\begin{equation}\label{eq:lw}
  3 h_t + \nabla_{x,z}\cdot(h^3\nabla_{x,z} \nabla_{x,z}^2h) 
  + \nabla_{x,z}\cdot\left[h^3\nabla_{x,z} K F(h)\right] = 0,
\end{equation}
where $\nabla_{x,z}$ stands for 2D in-plane gradient operator.
Equation~\eqref{eq:lw} is derived using the same scales as the ones used 
to obtain Eq.~\eqref{eq:NS_nondim}, with the same length scale chosen
for both the in--plane directions ($x$ and $z$) and the film thickness ($h$).
The LSA of Eq.~(\ref{eq:lw}) shows that
if the initial film thickness, $h_0$, is perturbed by a mode with infinitesimal
amplitude, $\delta$, of the form
$\exp(i(kx + lz))$, then the initial perturbation will grow or decay
with a growth rate
\begin{equation}\label{eq:growthrate}
  \beta = \frac{h_0^3(k^2+l^2)(k_c^2 - (k^2+l^2))}{3},
\end{equation}
where $k_c$ is the critical wavenumber, given by
\begin{equation}\label{eq:kc}
  k_c^2 = -\frac{K}{h_0}\left[ m\left(\frac{h^\ast}{h_0}\right)^m 
    - n\left(\frac{h^\ast}{h_0}\right)^n \right].
\end{equation}
Note that if $k_c^2 <0$, then $\beta<0$, and there is no instability for any wavenumber. 
If $k_c^2 > 0$, all modes with wavenumber $k<k_c$ are unstable.
Associated with Eq.~\eqref{eq:growthrate} is a wavenumber of maximum growth, $k_{\text{max}}$, and 
the corresponding maximum growth rate, $\beta_{\text{max}}$, given by
\begin{equation}\label{eq:kmax}
  k_{\text{max}} = \frac{k_c}{\sqrt{2}};\quad
  \beta_{\text{max}} = \beta(k^2+l^2 = k_{\text{max}}^2) = \frac{k_c^4}{12}. 
\end{equation}
We will frequently make reference to the wavelength of maximum growth, defined by
$\lambda_{\text{max}} = 2\pi/k_{\text{max}}$.

Within the L-W, these results imply the following features of the film instability:
\begin{itemize}
  \item The wavenumber of maximum growth, $k_{\text{max}}$, scales with $\sqrt{1-\cos\theta_{\text{eq}}}$. Large contact angles 
    imply larger $k_{\text{max}}$, and a corresponding decrease in $\lambda_{\text{max}}$.
  \item The maximum growth rate, $\beta_{\text{max}}$, scales with $(1-\cos\theta_{\text{eq}})^2$. Large contact angles dramatically
      increase the growth rate of the dominant mode, and thus reduce the time it takes for films to break up.
\end{itemize}

\section{Results}\label{sec:results_vdwreducedpressure}
\subsection{Contact Angles}\label{sec:contactAngles}

We first demonstrate that the method under consideration can yield
contact angles as accurately as the methods presented in~\cite{MahadyvdW15}.
For this purpose, we consider a 2D droplet, with the initial fraction (the
initial region occupied by the liquid phase, i.e. where $T=1$)
set to be the following
$$
  \{ (x,y) : x^2 + (y+R\cos\theta_i - h^\ast)^2 < R^2 \text{ or } y<h^\ast \}.
$$
Thus the initial profile is a circular cap sitting on top of an equilibrium
film such that it intersects this film with angle $\theta_i$.
The value of $R$ is chosen so that the total area of the circular cap is 
equal to $A_0 = 0.75^2\pi/2$.
For simplicity, for our first tests, we set $\theta_{\text{eq}} = \theta_i = \pi/2$.
Since for small $h^\ast$, $\theta_{\text{num}}$ is close to $\theta_{\text{eq}}$, the initial fluid
configuration is close to its equilibrium.
The droplet is then allowed to relax, and we measure 
$\theta_{\text{num}}$ via the method described in Sec.~\ref{sec:model}. 
We set $\mbox{Oh} = (0.05)^{-1/2}$, a sufficiently large value so that inertial 
effects are not significant. 

\begin{table}[t]
  \centering
  \begin{tabular}{l || c | c}
    $\Delta$ 	& R--P 	& B--F 	\\
    \hline
    $1/2^5$ 	& $6.1\times 10^{-2}$	& $7.0\times 10^{-3}$ 		\\
    $1/2^6$ 	& $2.6\times 10^{-3}$	& $4.3\times 10^{-3}$		\\
    $1/2^7$ 	& $8.0\times 10^{-4}$	& $3.8\times 10^{-4}$		\\
  \end{tabular}
  \caption[]{
  Comparison of the mesh convergence between the R--P
    method  and the B--F method of~\cite{MahadyvdW15} for a drop.  
Here, $\Delta$ is the grid resolution used (the maximum one for the R--P method, and uniform
for the B--F method), and the L$^1$ norm of the error in the profile 
shape.  The performance of the two methods is comparable at higher resolutions. }
    \label{tab:err_meth}
\end{table}
Table~\ref{tab:err_meth} compares the convergence of the equilibrium droplet
profile as a function of the minimum mesh size, for a droplet simulated
using the R--P method discussed here, and the B--F method of~\cite{MahadyvdW15}.
The R--P simulations are resolved to a resolution $\Delta_{\text{max}}=\Delta$ 
on an adaptive mesh as shown in Fig.~\ref{fig:mesh}, while 
the B--F method simulations are refined uniformly with mesh size $\Delta$.
The error is measured as the L$^1$ norm of the error in the profile shape.
The equilibrium film thickness is $h^\ast=0.03$.
Despite the fact that the B--F simulations are run with a uniform mesh,
the R--P method performs comparably well in 
convergence in mesh to the B--F method at higher resolutions.

\begin{figure}[t]
  \centering
  \includegraphics[width=3in]{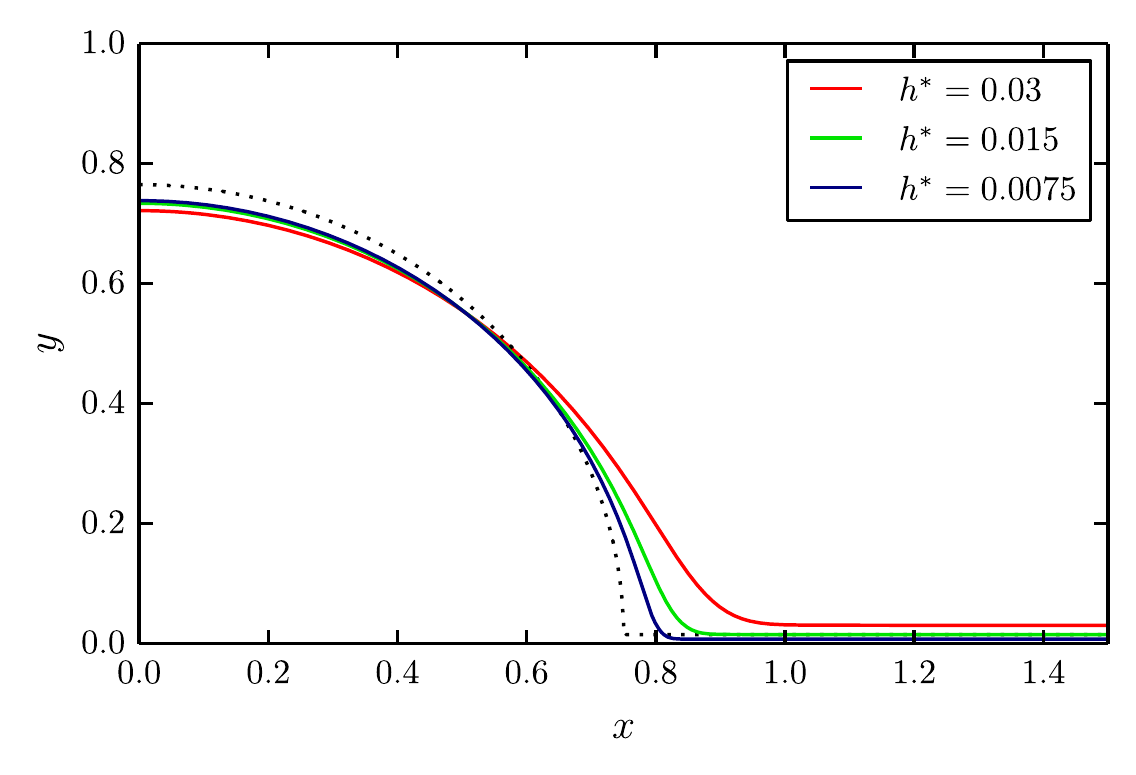}
  \caption[]{Effect of $h^\ast$ on the profile when $\theta_i = \theta_{\text{eq}} = \pi/2$. 
	The dotted line shows the initial profile for $h^\ast=0.015$.}
  \label{fig:conv_hstar}
\end{figure}

Figure~\ref{fig:conv_hstar} shows simulation profiles obtained using
the R--P method, again with $\theta_{\text{eq}}=\theta_i=\pi/2$.
The profiles at equilibrium are shown for $h^\ast=0.03$ (red) , $h^\ast=0.015$ (green),
and $h^\ast=0.0075$ (blue).
The initial condition for $h^\ast=0.015$ is plotted by the black dotted line,
showing the profile of a droplet with contact angle $\theta_{\text{eq}}$.
The smooth transition from the droplet profile to the equilibrium film is 
clearly visible; as $h^\ast$ is reduced, the equilibrium profiles approach
those of a droplet with contact angle $\theta_{\text{eq}}$.

\begin{table}[t]
  \centering
  \begin{tabular}{l || c  | c }
    $h^\ast$  	& R--P  	 	& B--F \\
    \hline
    0.03	& 1.36 		& 1.37 	\\
    0.015	& 1.45		& 1.47 \\
    0.0075 	& 1.49 		& 1.53
  \end{tabular}
  \caption[]{The contact angle, $\theta_{\text{num}}$, obtained by the R--P and the B--F methods for 
  different values of $h^\ast$.    The contact angles approach
  $\theta_{\text{eq}} = \pi/2$ at a similar rate.}
  \label{tab:hstar}
\end{table}
Table~\ref{tab:hstar} shows the measured values of $\theta_{\text{num}}$,
again for the droplet with $\theta_{\text{eq}} = \theta_i = \pi/2$, 
as $h^\ast$ is varied, for both the B--F and the R--P methods.
While the values differ slightly between the two methods, as $h^\ast$ is reduced,
$\theta_{\text{num}}$ becomes closer to $\theta_{\text{eq}}$. 

We conclude that the R--P method is comparable to the B--F method
in its ability to model the contact angle. 
In addition, it comes with the potential for dramatically improved performance,
due to fact that the layer of fluid near the substrate does not 
necessarily need to be highly resolved. For example,
for the drops simulated in Table~\ref{tab:err_meth}, with the timestep fixed at $10^{-5}$, 
at $\Delta=1/2^6$ the R--P method takes approximately $14\%$
as much CPU time as the B--F method; at $\Delta=1/2^7$, the R--P
method has approximately $5\%$ the runtime.
This is because, for each timestep, the number of computations each method incurs is
$O(N)$, where $N$ is the number of cells. 
However, the adaptive mesh illustrated in Fig.~\ref{fig:mesh} has approximately $O(1/\Delta)$ cells,
while a uniform mesh has $O(1/\Delta^2)$ cells. 
Simple adaptive meshes were used in~\cite{MahadyvdW15}, improving the performance
of the B--F method, however, higher resolution is still required at a layer of nonzero 
thickness near the substrate. The complexity difference is similar in 3D, where the R--P
method scales as $O(1/\Delta^2)$ and the B--F method as $O(1/\Delta^3)$.

The improvement in performance permits the study of problems that are impractical to 
consider using the B--F method, including breakup of the films in 2D and 3D.  We proceed
to analyze these problems.  

\subsection{Film Instability: Linear regime}\label{sec:linearinstability}

In this section, we compare the LSA predictions of the L-W equation
with the results of simulations, both applied to the instability of thin films in 2D.
We focus here on early times/linear regime for which
the LSA is expected to hold and discuss the comparison between the results influenced
by the relevant parameters.   In particular 
we focus on the influence of the unperturbed film thickness, $h_0$, and of
the equilibrium contact angle, $\theta_{\text{eq}}$.  
Note that the LSA should 
apply even for large contact angles, since we focus on early stages of instability 
when the interfacial slopes are small.
We set $\mbox{Oh}=0.45$, which implies a low inertia regime, so that the assumptions
of the L-W are reasonably well satisfied (in Sec.~\ref{sec:2D}, we 
discuss the values of $\mbox{Oh}$ that constitute low inertia in the present context).
For all the simulations, the initial region occupied
by the liquid phase (i.e., $T=1$) is between $y=0$ and
$h=h(t=0,x) = h_0(1 +  \delta\cos(kx))$, where $\delta=0.08$. 
For all simulations in this section, we set $h^\ast=0.12$.

\begin{figure}[t]
    \includegraphics[width=0.9\textwidth]{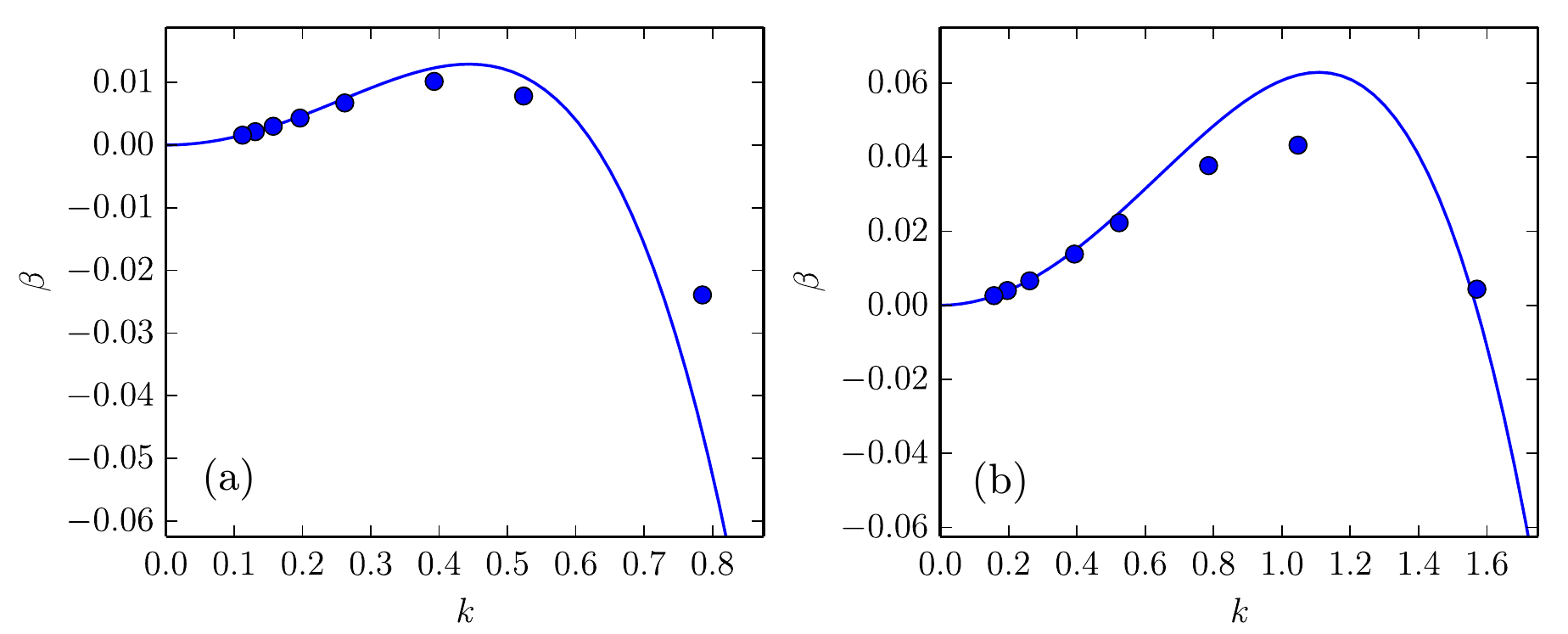}
  \caption[]{ Comparison between the 
        dispersion curve predicted by the LSA of the
    	L-W equation (blue solid curve) and the growth rate computed by R--P  based simulations (symbols), for
      	$\theta_{\text{eq}}=\pi/2$, for (a) $h_0=1.0$ and (b) $h_0=0.5$.  
	}
  \label{fig:LSA}
\end{figure}

Figure~\ref{fig:LSA}  shows the growth rates resulting from the LSA together
with the ones extracted from simulations for early time evolution,  for a discrete 
set of wavenumbers.
We measure the growth rate of simulations according to the following procedure:
first, we output the minimum $y$-coordinate of the fluid interface, $h_{min}(t)$.
We then find $t_0$ and $t_1$, such that for $t_0 < t < t_1$, the function
$\log(h_{min}(t))$ is approximately linear in $t$; $t_0$ is generally near $0$,
and $t_1$ is small, so this corresponds to the interval of time
when the growth of the unstable modes is governed by the LSA.
We then perform a least-squares
fit of a line to $\log(h_{min}(t))$ over this interval. 
Repeating the same procedure for the maximum $y$-coordinate of the fluid interface,
$h_{max}(t)$, we calculate the growth rate as the average of the slopes of 
the fitted lines.

We focus on the influence of $h_0$, and set $\theta_{\text{eq}} = {\pi/2}$.  
For small wavenumbers, $k$, the agreement is very good; however, 
for larger values of $k$, there are significant differences, for both values of $h_0$.   
We conjecture that the differences are due to the fact that the L-W
assumptions no longer hold for larger values of $k$, since the separation of scales required
for the L-W  to apply may not be satisfied.  For example,
if one considers the relevant lengthscales as   $x_c= \lambda_{\text{max}}$ and $h_c=h_0$, in 
the in-plane and out-of-the-plane direction, respectively,  
then for $h_0 = 0.5$ one finds $\epsilon  = x_c/h_c$ is $O(0.1)$ and therefore
for  $k$ values larger than $k_{\text{max}}$, the requirement $\epsilon \ll 1$ is no longer a good approximation.   
However, it should be noted that the differences between the simulations and the LSA  are mostly focused 
on large values of $k$, where films are stable.  

One may expect that the LSA would lead to more accurate results 
for thinner films.  As we can see in Fig.~\ref{fig:LSA}, this is not the case.  An insight can
be reached again by considering the relevant length scales; the ratio of the film thickness, $h_0$, and the 
relevant in-plane length scale, $\lambda_{\text{max}}$, given by 
\begin{equation}\label{eq:smallParameter}
 \frac{h_c}{\lambda_{\text{max}}} = \frac{1}{2\pi}\sqrt{-Kh_0
  \left[ m\left(\frac{h^\ast}{h_0}\right)^m - n\left(\frac{h^\ast}{h_0}\right)^n \right]}.
\end{equation}
This ratio is a decreasing function of $h_0$ for 
$
  h_0 > h^\ast\left[m(m-1)/(n(n-1))\right]^{m-n}.
$  Therefore, reducing $h_0$ actually implies that the LSA is less accurate, except
when $h_0$ is nearly as small as $h^\ast$.

\begin{figure}[h!]
  \centering
    \includegraphics[width=\textwidth]{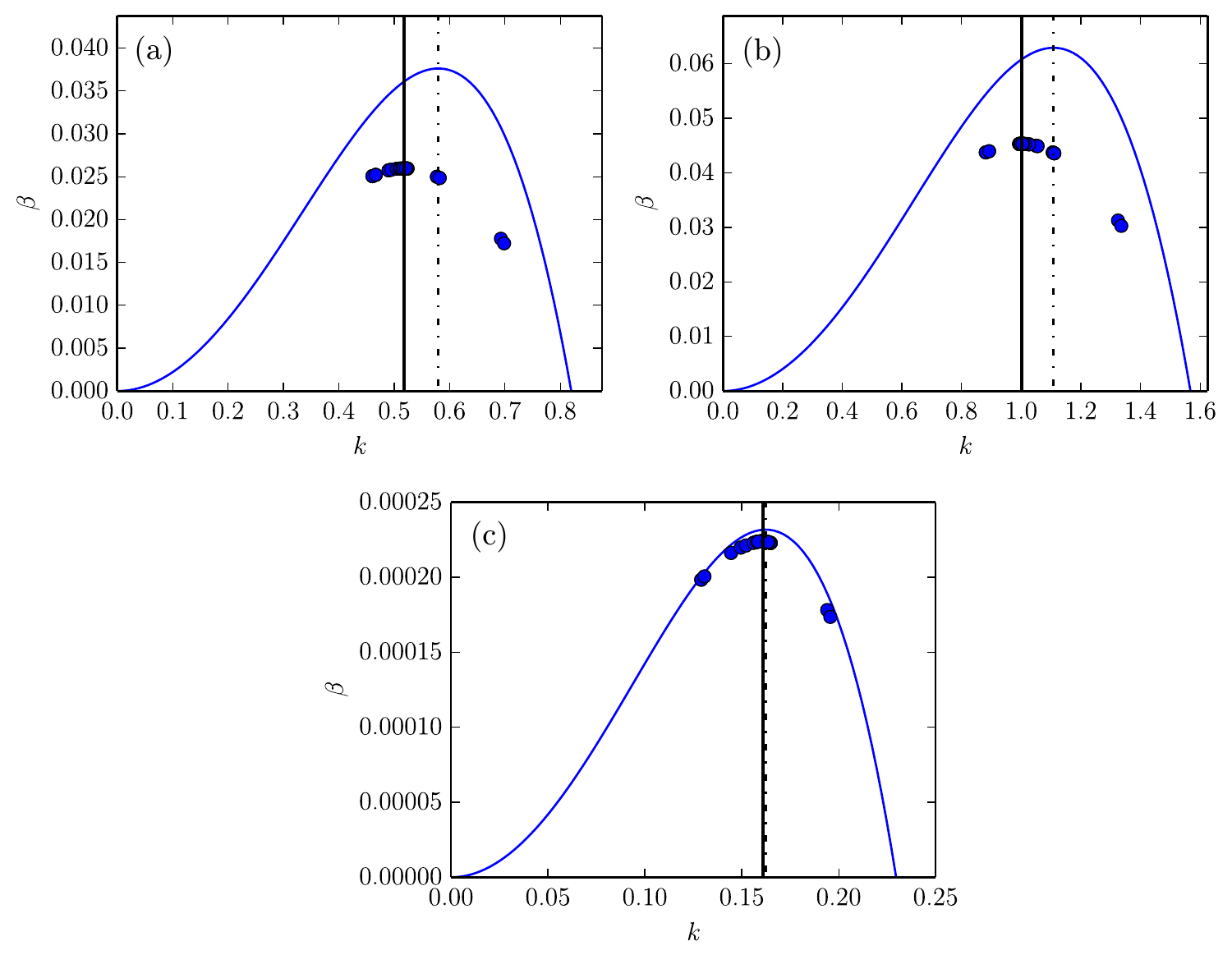}
     \caption[]{
        Comparison between the growth rate predicted by the LSA (blue solid curve),
	and the simulations (symbols), for $h_0=0.5$, for
	(a) $\theta_{\text{eq}} = 3\pi/4$, 
	(b) $\theta_{\text{eq}} = \pi/2$, and 
	(c) $\theta_{\text{eq}} = \pi/6$. 
	The value of $k_{\text{max}}$ obtained from simulations is
	shown by the vertical solid line; the dash-dotted line shows
	the value of $k_{\text{max}}$ predicted by the LSA. } 
	\label{fig:LSAtheta}
\end{figure}
Equation~\eqref{eq:smallParameter} implies that a reduction in $\theta_{\text{eq}}$ should
improve agreement with the predictions of the LSA. 
Figure~\ref{fig:LSAtheta} shows the results for three values of $\theta_{\text{eq}}$ and we
indeed immediately observe much smaller differences between the simulations and the
LSA for smaller $\theta_{\text{eq}}$.   For 
larger $\theta_{\text{eq}}$, the LSA strongly overpredicts the growth rates as well as the 
values of $k_{\text{max}}$.     We will see in the next section that the differences between
the values of $k_{\text{max}}$ resulting from the simulations and the LSA become 
much more visible in the nonlinear regime of instability development.

\subsection{Film instability: Nonlinear evolution and breakup - 2D}
\label{sec:2D}

Here we focus on the nonlinear process of film breakup, and concentrate in
particular on the role of the contact angle, $\theta_{\text{eq}}$, and
Ohnesorge number, $\mbox{Oh}$, on the properties of emerging patterns. 
While the discussion that follows is general, and is formulated
in terms of previously defined nondimensional quantities, in order
to choose a reference parameter set, we consider nanoscale liquid
metallic films. The stability properties of such
films have been a subject of recent interest in connection
with nanoparticle fabrication (see
e.g.~\cite{ajaev_pof03,trice_prl08,kd_pre09,fuentes_pre11,khenner_pof11,
Fowlkes2013}). In experiments, a film of metal, with thickness on 
the order of tens of nanometers, is deposited on a surface.
This film is then liquefied, and subsequently 
breaks up into droplets, which solidify to form nanoparticles.

For definitiveness, we focus on Cu films, as considered in~\cite{gonzalez2013}.
The viscosity, density, and surface tension are taken to be the values of
liquid Cu at its melting point~\cite{gonzalez2013}, with $\rho=7760$ kg/m$^3$,
$\mu = 0.00438$ Pa$\cdot$s, and $\gamma = 1.304$ N/m.  We set relevant length
scale, $L$, as  the reference film thickness, taken to be $8$ nm.
The Ohnesorge number corresponding to these parameters is 
	$\mbox{Oh} = 0.487$.  
	Note that this represents a rough lower bound for 
	$\mbox{Oh}$ in the motivating experiments: 
	temperatures may exceed the melting point substantially,
      reducing the viscosity, and furthermore the relevant 
	length scales may be larger than $8$ nm.
	This variation in
$\mbox{Oh}$ makes it important to explore the boundary between the viscous and
inertial regimes.
The simulations that we present in this section are for a
film of thickness $h_0=1$; we also considered films with $h_0=0.5$, and
obtained similar results.  We use $h^\ast=0.1225$, and $\theta_{\text{eq}} =
79^\circ \approx 0.439\pi$.  Note that for these parameters,
$\lambda_{\text{max}} = 15.6$, and $\lambda_c =11.0$.

\begin{figure}[t]
  \centering
     \includegraphics[width=0.9\textwidth]{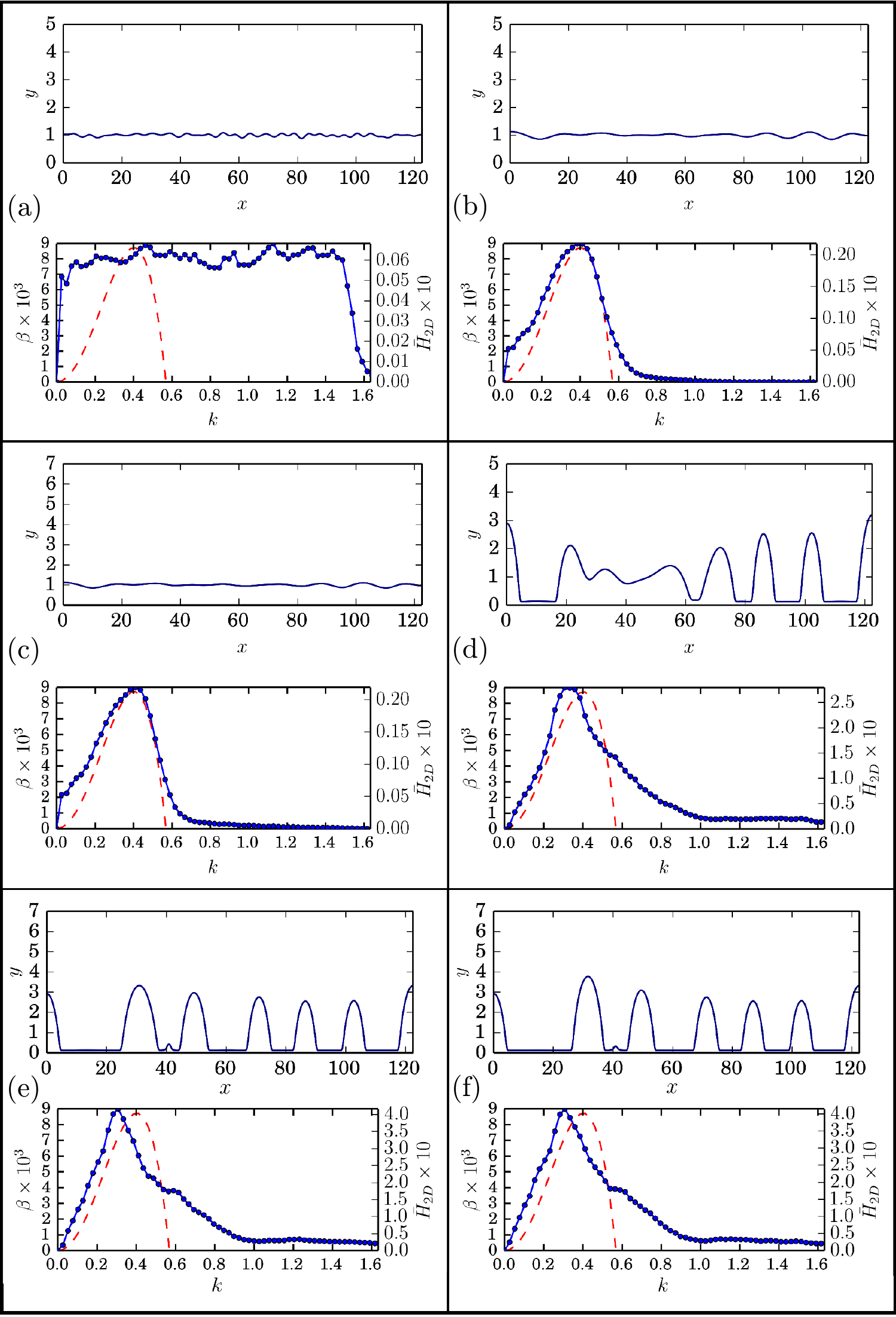}
  \caption[]{Time evolution of a representative simulation of film breakup in 2D
    with $h_0=1$, $\mbox{Oh} = 0.487$, and $\theta_{\text{eq}}=0.439\pi$;
    	the solid blue curve shows the fluid interface. 
  	The associated Fourier spectrum, averaged over 20 realizations, is shown
	below each image. (a) $t=0$, (b) $t=179$, (c) $t=357$, (d) $t=536$, (e) $t=715$, and (f) $t=882$.}
  \label{fig:2D_spectra}
\end{figure}
\begin{figure}[h!]
  \centering
       \includegraphics[width=0.9\textwidth]{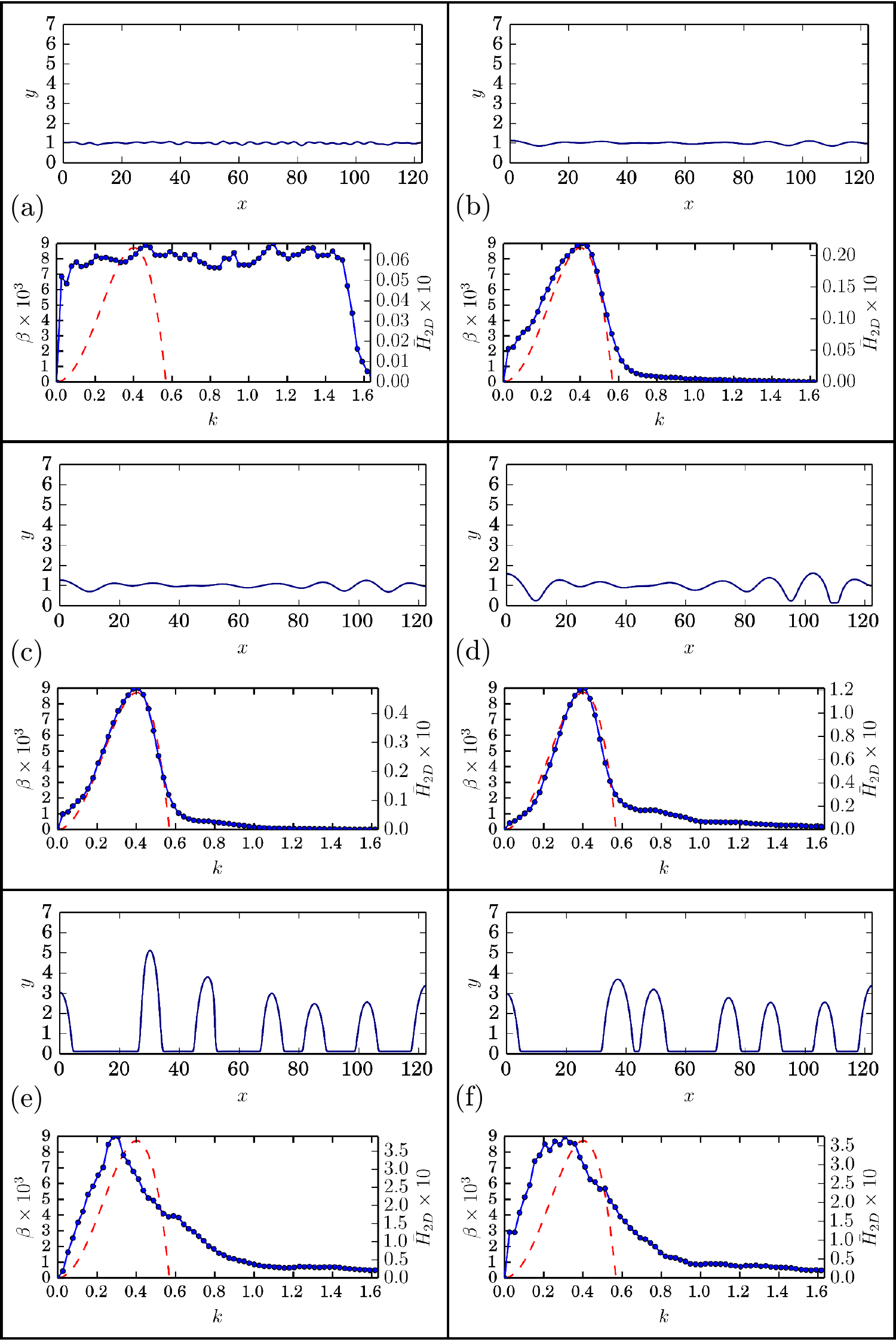}
    \caption[]{Time evolution of a representative simulation of film breakup in 2D
    with $h_0=1$, $\mbox{Oh} = 0.0487$, and $\theta_{\text{eq}}=0.439\pi$;
    	the solid blue curve shows the fluid interface. 
  	The associated Fourier spectrum, averaged over 20 realizations, is shown
	below each image. (a) $t=0$, (b) $t=357$, (c) $t=536$, (d) $t=715$, (e) $t=1429$, and (f) $t=2858$.}
  \label{fig:2D_spectra_0p1Oh}
\end{figure}

For 2D simulations, the computational domain is specified by $(L_x,L_y)$, with $L_x \approx 8 \lambda_{\text{max}}$, and 
$L_y$ sufficiently large so that the value assigned to it does not influence the results. For the simulations discussed
next, we use $L_x = 122.5$, $L_y = 43.125$, 
and the initial film shape specified by 
is $h_0 + \zeta(x)$, where
\begin{equation}\label{eq:perturbation}
  \zeta(x) = \sum_{i=1}^{60} \delta_i \cos\left(\frac{2\pi x}{\lambda_i}\right).
\end{equation}
Here, $\lambda_i = 2L_x/i$, and $\delta_i$ is a random perturbation amplitude,
uniformly distributed in the range $\pm 0.0125$.
We simulate the evolution with $N_s = 20$ sets of $\delta_i$.
For each simulation (numbered $j$), a discrete height profile is produced,
$\hat{h}_j(t,x)$.
We then compute the discrete Fourier transform (DFT) of each height profile,
$\hat{H}_j(t,k)$. Finally, we compute the average of these as
\begin{equation}\label{eq:avgFourier2D}
  \bar{H}_{2D}(t,k) = \frac{1}{N_s}\sum_{j=1}^{N_s} \hat{H}_j(t,k).
\end{equation}
Figure~\ref{fig:2D_spectra} shows the time evolution of a typical profile
$\hat{h}_1(t,x)$ (the top image in each of Fig.~\ref{fig:2D_spectra}(a)~-~(f)), 
$\bar{H}_{2D}(t,k)$ along with the dispersion curve from the LSA
(the bottom image in each of Fig.~\ref{fig:2D_spectra}(a)~-~(f), 
smoothed with a running $5$ point average).  
We see that as the initial perturbations
begin to grow, the $\bar{H}_{2D}(t,k)$ attains the form similar to the dispersion
curve (Fig.~\ref{fig:2D_spectra}(a)~-~ (c)).  However, as the holes and consequently drops
start to form, the peak in
$\bar{H}_{2D}(t,k)$ shifts towards smaller values of $k$, so that the final
distribution of drops is characterized by a length scale larger than
$\lambda_{\text{max}}$.
\clearpage
Next we consider different values of $\mbox{Oh}$.
For $\mbox{Oh}=4.87$, no significant difference is observed compared to $\mbox{Oh} = 0.487$, indicating
that both sets of results belong to the high viscosity limit.
Smaller values of $\mbox{Oh}$, however, lead to different results.
Figure~\ref{fig:2D_spectra_0p1Oh} shows profiles and associated averaged
Fourier spectra for a film with $\mbox{Oh}=0.0487$.   We see that the 
film takes much longer to 
break up compared to the films characterized by larger $\mbox{Oh}$.   Also, 
the long term evolution of the Fourier spectrum shows a flatter peak,
when compared with the larger $\mbox{Oh}$ simulations, indicating that the preference for a 
specific wavenumber is weaker in the late stages of evolution.   The practical consequence
of this may be increased degree of disorder in the distribution of emerging patterns.

\begin{figure}[t]
  \centering
  \includegraphics{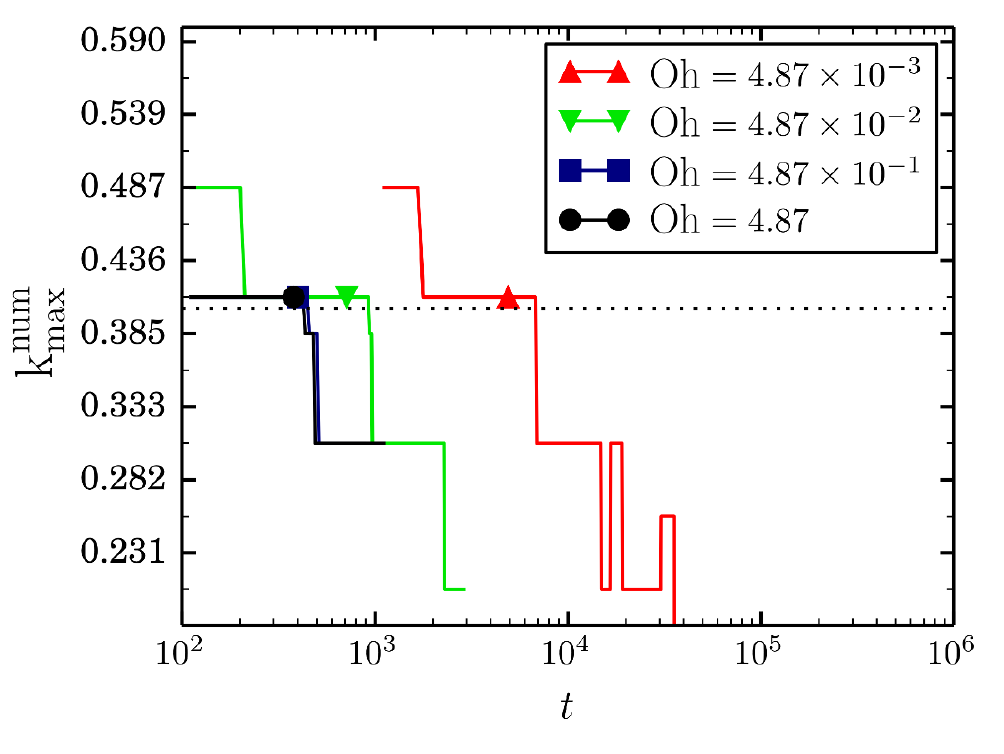}
  \caption{ 
  Comparison of $k_{\text{max}}^{\text{num}}$ as a function of time for films of
  varying $\mbox{Oh}$.  Symbols show the approximate breakup times for each parameter
  set, and the solid dashed line shows $k_{\text{max}}$ predicted by the LSA. 
  Note that $\mbox{Oh}=4.87$ and $\mbox{Oh}=0.487$ are visually nearly indistinguishable. 
  The ticks on the $y$-axis are for wavenumbers that can be resolved on the finite domain.
  Here, $h_0=1$, $\theta_{\text{eq}}=0.439\pi$.}
  \label{fig:kmaxes_2D}
\end{figure}

In order to study the effect of $\mbox{Oh}$ on the evolution of the film more systematically,
we define $k^{\text{num}}_{\text{max}}(t)$ as the value of $k$ for which $\bar{H}_{2D}(t)$
attains its maximum. 
Figure~\ref{fig:kmaxes_2D} plots $k^{\text{num}}_{\text{max}}$
for various $\mbox{Oh}$ as a function of time (note that due to the 
finite domain size, $k_{\text{max}}^{\text{num}}$ only takes discrete values, leading
to the staircase appearance of these plots).
The symbols on each curve indicate the approximate time at which the film ruptures,
and the dashed line shows the value of $k_{\text{max}}$ from the LSA.
The overall behavior is similar for all curves: $k_{\text{max}}^{\text{num}}$ quickly
takes a value as close to the $k_{\text{max}}$ as can be resolved on a finite domain, and then
stays at this value until the film breaks up. Afterwards,
it decreases, indicating that the length scale of the final droplet
distribution is larger than $\lambda_{\text{max}}$.    We note that 
for $\mbox{Oh}=4.87$ and $\mbox{Oh}=0.487$, the behavior is nearly indistinguishable, indicating
large $\mbox{Oh}$ limit.
Most importantly, smaller values of $\mbox{Oh}$ significantly increase the time it
takes for the film to rupture.
The LSA does not capture this behavior, and instead predicts 
that the breakup time is approximately  the same ($\approx 500$)
for all values of $\mbox{Oh}$, consistently with the results of large $\mbox{Oh}$ simulations. 

\begin{figure}[!th]
  \centering
  \includegraphics[width=4in]{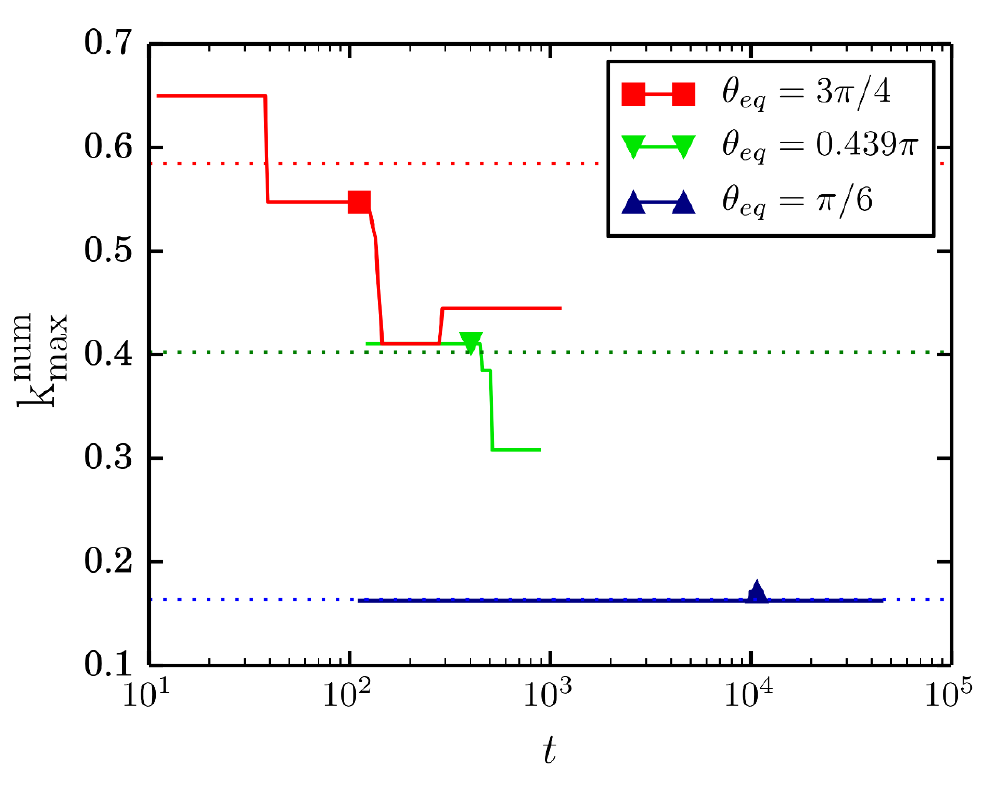}
  \caption{Plot of $k_{\text{max}}^{\text{num}}$ for varying $\theta_{\text{eq}}$. The green curve
	corresponds to 	$\mbox{Oh}=0.487$ in Fig.~\ref{fig:kmaxes_2D}. The symbols show
	the approximate time of film rupture. Each dashed line shows $k_{\text{max}}$ from
    	the LSA for the curve of the same color.}
  \label{fig:kmax_theta}
\end{figure}

Next, we investigate the effect of the contact angle on the rupture.
We carry out two additional simulation sets, both with $\mbox{Oh}=0.487$, and with
different values of $\theta_{\text{eq}}$ and different domain sizes, keeping 
$L_x \approx 8 \lambda_{\text{max}}$:
\begin{enumerate}
  \item $\theta_{\text{eq}}=\pi/6$, $L_x = 367.5$; 
  \item $\theta_{\text{eq}}=3\pi/4$, $L_x = 91.875$.
\end{enumerate}
For each set, the form of the perturbation is the same as in Eq.~\eqref{eq:perturbation},
using corresponding values of $L_x$.  Figure~\ref{fig:kmax_theta} shows 
$k_{\text{max}}^{\text{num}}$ for these two contact angles, as well as for 
$\theta_{\text{eq}}=0.439\pi$. 
For each curve, the symbol of the same color indicates the breakup time, and the 
dashed line of the same color shows $k_{\text{max}}$ from the LSA.
Note that the breakup times predicted by the LSA are $\approx 18000$, $500$,
and $110$, for $\theta_{\text{eq}}=\pi/6$, $0.439\pi$, and $3\pi/4$, respectively,
roughly in line with the simulated breakup time for all cases. 
For $\theta_{\text{eq}}=3\pi/4$, the behavior is similar to $\theta_{\text{eq}}=0.439\pi$. However,
the evolution of $\theta_{\text{eq}}=\pi/6$ is dramatically simpler: almost immediately,
$k_{\text{max}}^{\text{num}}$ relaxes to near $k_{\text{max}}$, and remains at approximately the same value
for the considered simulation times. 

To gain a basic understanding of the influence of $\theta_{\text{eq}}$ on the evolution, 
consider the following simple argument.  Approximate the profile of a rupturing film 
of unit initial thickness, perturbed by a mode of wavelength $\lambda_{\text{max}}$, by 
$h(x) = 1 + A(t) \cos\left(2\pi x/\lambda_{\text{max}}\right),$
where $A(t)$ is the time dependent amplitude. At the time of breakup,
$A\approx 1$.  Then, further approximate the instantaneous contact angle, $\theta_r$, of the rupturing
film by considering the slope 
at the point of inflection, given by
$\tan\theta_{r} =2\pi/\lambda_{\text{max}}$.
Substituting the value of  $\lambda_{\text{max}}$, we obtain
$\tan\theta_{r} = B\sqrt{1-\cos\theta_{\text{eq}}},$
with $B= \sqrt{-\left(m{h^\ast}^m - n{h^\ast}^n\right)2\sqrt{2}h^\ast}.$
Consider now the difference between $\theta_r$ and $\theta_{\text{eq}}$.   This 
difference is an increasing function of $\theta_{\text{eq}}$, 
suggesting that when films with larger $\theta_{\text{eq}}$ rupture, the corresponding
$\theta_r$ is significantly different from $\theta_{\text{eq}}$, leading to quick retraction
of the film from the point of rupture.  This retraction reduces the space available for 
consecutive breakups, and leads to larger distances between the drops than predicted
by the LSA.  For smaller values of $\theta_{\text{eq}}$, the difference between $\theta_r$  and 
$\theta_{\text{eq}}$ is small at the point of initial rupture, and the retraction effect
is not present.   

Next we comment on the difference between $k_{max}^{num}$ and the corresponding $k_{max}$, visible in
Figs.~\ref{fig:kmaxes_2D} - \ref{fig:kmax_theta}.  First we note that this 
shift is not affected by inertial effects for the simulated times:  Fig.~\ref{fig:kmaxes_2D} 
shows that $\mbox{Oh}=48.7$ and $\mbox{Oh}=0.0487$  lead to  the same $k_{max}^{num}$ after breakup (the additional
coarsening expected for very long times is beyond the scope of the present study).  
These results cannot be explained by the breakdown of the LSA for large $\theta_{eq}$: from Fig.~\ref{fig:LSAtheta},
we might expect a roughly 10\% difference between the wavelength of maximum growth in LSA
and simulations, which is significantly less than the difference between  $k_{max}^{num}$ and 
$k_{max}$ after the film has ruptured.  
Thus, we expect that the difference is due to nonlinear effects relevant during breakup for
larger contact angles.  We note here that for for smaller contact angles 
the effect is not  visible within the present degree of accuracy; the simulations carried out in larger
domains and with larger number of realizations
using long-wave approach suggest that (much less obvious) coarsening effects may be seen within long-wave simulations as well~\cite{nesic_15a}.

\begin{figure}[h!]
  \centering
    \centering
    \includegraphics[width=\textwidth]{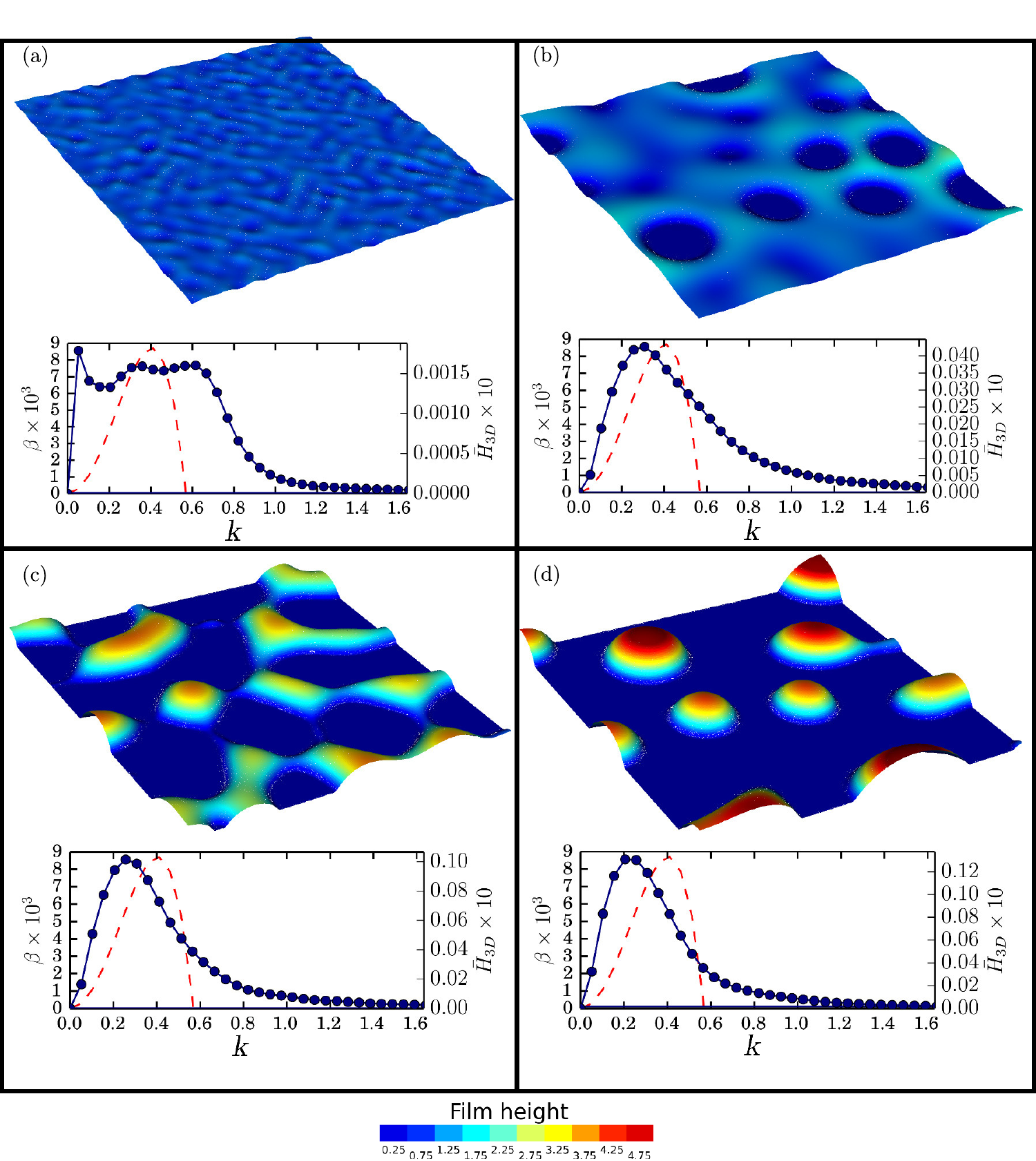}
  \caption[]{Time evolution of a representative simulation of film breakup in 3D,
    	with $\mbox{Oh}=0.487$, $h_0=1.0$, and $\theta_{\text{eq}}=0.439\pi$.
    	The color shows the logarithm of the height of the interface above the
	substrate.
  	The associated Fourier spectrum is shown
	below each image; these data are averaged over 10 instances, and
	smoothed with a 5 point running average.
	(a) $t=0.0$, (b) $t=335$, (c) $t=558$, and (d) $t=781$.}
  \label{fig:3D_spectra}
\end{figure}
\subsection{Film instability: Nonlinear evolution and breakup - 3D}
\label{sec:3D}
\begin{figure}[t]
  \centering
  \includegraphics{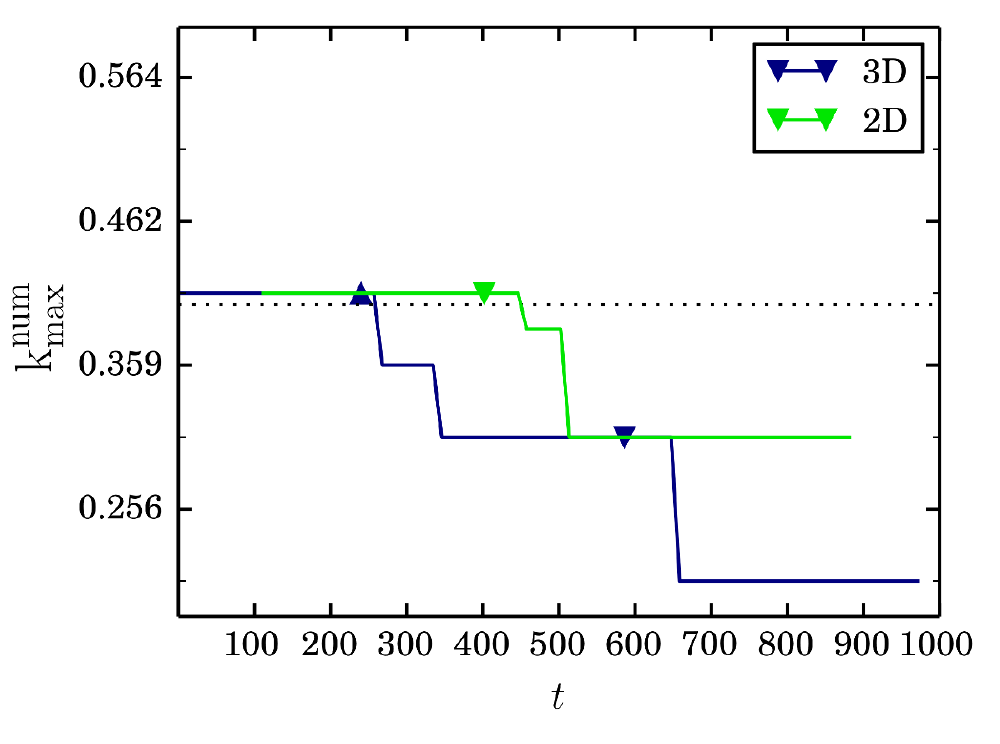}
  \caption{Comparison of $k_{\text{max}}^{\text{num}}$ as a function of time for
  3D films (blue) and 2D films (green).  The green inverted triangle shows the approximate breakup
  time of the 2D simulation. The blue triangle shows the approximate time at which the first
  holes form in the 3D film; the inverted blue triangle shows the approximate time that droplets begin
  to form. The 
  solid dashed line shows $k_{\text{max}}$ predicted by the LSA.}
  \label{fig:kmax_3D}
\end{figure}

Next, we consider film breakup in 3D. The setup is identical to the 2D case, except that the domain
size is specified by $L_x = L_z = 4 \lambda_{\text{max}}$ and $L_y  = 31.25$.  
The initial condition is a perturbed film specified by  $h_0 + \zeta(x,z)$, and 
\begin{equation}\label{eq:3d_IC}
  \zeta(x,y) = \sum_{i=1}^{30} \sum_{j=1}^{30} \delta_{ij} 
  	\cos\left( \frac{2\pi x}{\lambda_i} \right)
	\cos\left( \frac{2\pi z}{\lambda_j} \right),
\end{equation}
where $\lambda_i=2L_x/i$, and $\delta_{ij}$ are random perturbation amplitudes, uniformly
distributed in the range $\pm 0.0125$.
We simulate the film for $N_s = 10$ sets of $\delta_{ij}$. 
As in the 2D case, we produce a height profile for each simulation, $\hat{h}_j(t,x,z)$, compute its
discrete Fourier transform, $\hat{H}_j(t,k,l)$, and the average
\begin{equation}\label{eq:avgFourier3D}
  \bar{H}_{3D}(t,k) = \frac{1}{N_s}\sum_{j=1}^{N_s} \frac{1}{2}\left( \hat{H}_j(t,k,0) + \hat{H}_j(t,0,k)\right).
\end{equation}
The summand in Eq.~\eqref{eq:avgFourier3D} represents the average of the discrete Fourier transforms along the coordinate
axes, taking advantage of the symmetry of Eq.~\eqref{eq:growthrate} to get more information about the growth rates.

Figure~\ref{fig:3D_spectra} shows the time evolution of $\hat{h}_1(t,x,z)$ (top
image in each part of the figure), and  $\bar{H}_{3D}$ along with the dispersion curve
predicted by Eq.~\eqref{eq:growthrate} (bottom images).
The plots of $\bar{H}_{3D}$ have been smoothed by a $5$ point running average. 
A similar trend is observed as in Fig.~\ref{fig:2D_spectra}: the spectrum takes on a similar
profile to the dispersion curve from the LSA, and as breakup proceeds, its peak, $k_{\text{max}}^{\text{num}}$,
shifts towards smaller wavenumbers.

Figure~\ref{fig:kmax_3D} shows $k_{\text{max}}^{\text{num}}(t)$ for both 2D and 3D films ($\mbox{Oh}=4.87$). 
For both 2D and 3D, $k_{\text{max}}^{\text{num}}$ is close to $k_{\text{max}}$ until the film ruptures, after which
both $k_{\text{max}}^{\text{num}}$ relax to smaller values.  However, once
droplets begin to form in 3D simulations,
$k_{\text{max}}^{\text{num}}$ relaxes to an even smaller value.   The same retraction argument that we used
to explain the evolution of $k_{\text{max}}^{\text{num}}(t)$ in 2D can as well be used here to explain this phenomenon.

\begin{figure}[t]
	\includegraphics{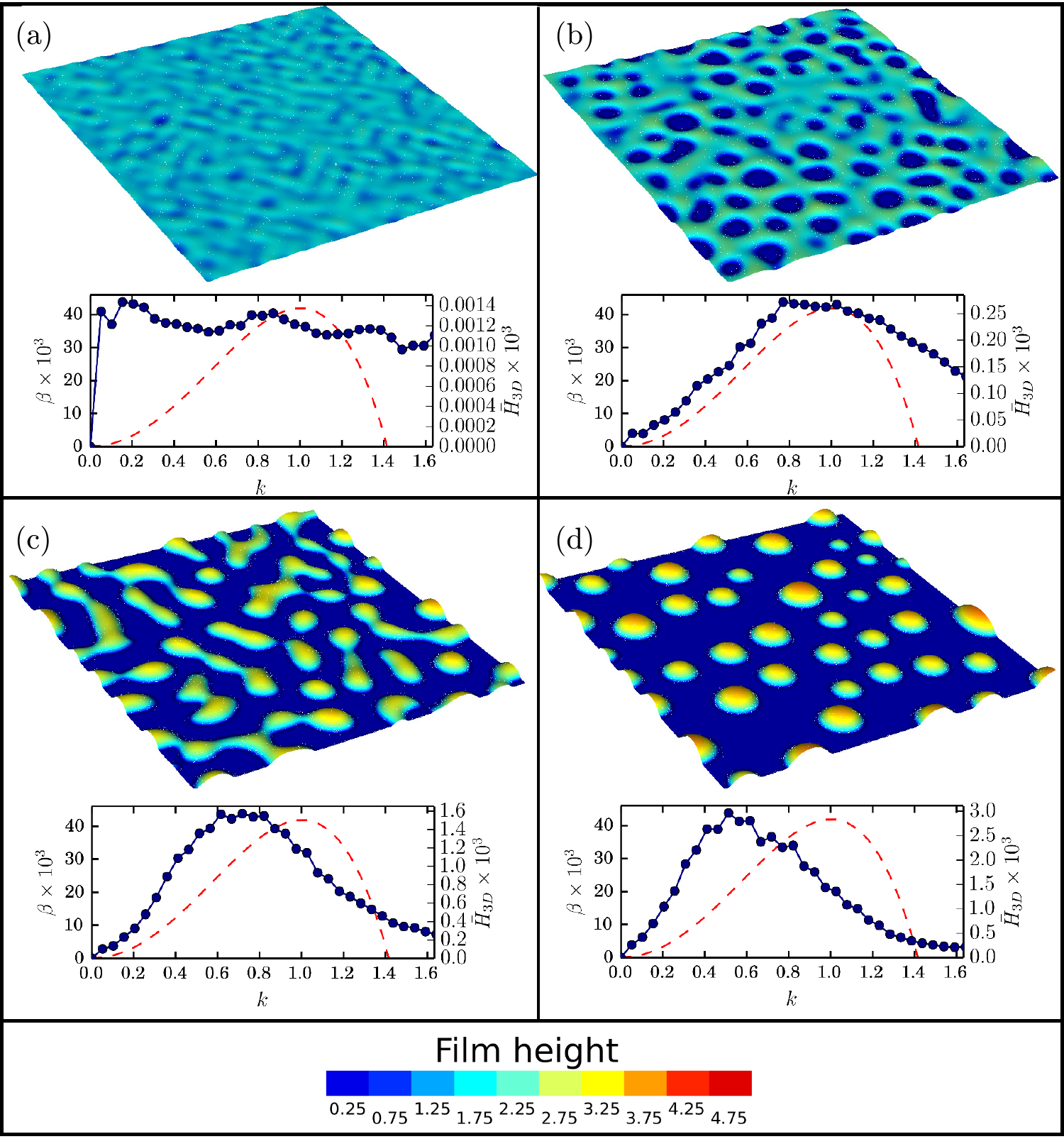}
	\caption[]{
		Time evolution of a film breakup
		in 3D, with $h_0=0.5$ and all parameters otherwise
		identical to the simulation presented in
		Fig.~\ref{fig:3D_spectra}.  The color shows the logarithm
		of the height of the interface above the substrate.  The
		associated Fourier spectrum is shown below each image; these
		data are averaged over 10 instances, and smoothed with a 5
		point running average:
		(a) $t=0.0$, (b) $t=45$, (c) $t=134$, and (d) $t=301$.}
	\label{fig:4nmevolution}
\end{figure}

We also study the effect of
initial film thickness on the evolution of the Fourier spectra for 3D films.
Figure~\ref{fig:4nmevolution} presents the
evolution of the film with an initial thickness $h_0=0.5$, while
keeping $\theta_{eq}$, $h^\ast$, and $\mbox{Oh}$ the same as in
Fig.~\ref{fig:3D_spectra}.
Our initial condition is again subject to a
white noise perturbation (Fig.~\ref{fig:4nmevolution}(a)). 
Holes form rapidly (Fig.~\ref{fig:4nmevolution}(b)), due to
the larger growth rate associated with the smaller $h_0$, when compared
with Fig.~\ref{fig:3D_spectra}.
$\bar{H}_{3D}$ shows a
peak near $k_{max}$ at the time of the formation of holes; however, this does
not agree as closely with the overall shape of the dispersion curve as is
observed for thicker films.  We conjecture that this difference is due to the
fact that holes form almost immediately, so that there is insufficient time for
$\bar{H}_{3D}$ to relax to the shape of the dispersion curve.  Similarly to
thicker films, as drops begin to form, $k_{max}^{num}$ relaxes to wavenumbers
significantly smaller than $k_{max}$.

\begin{figure}[t]
	\includegraphics{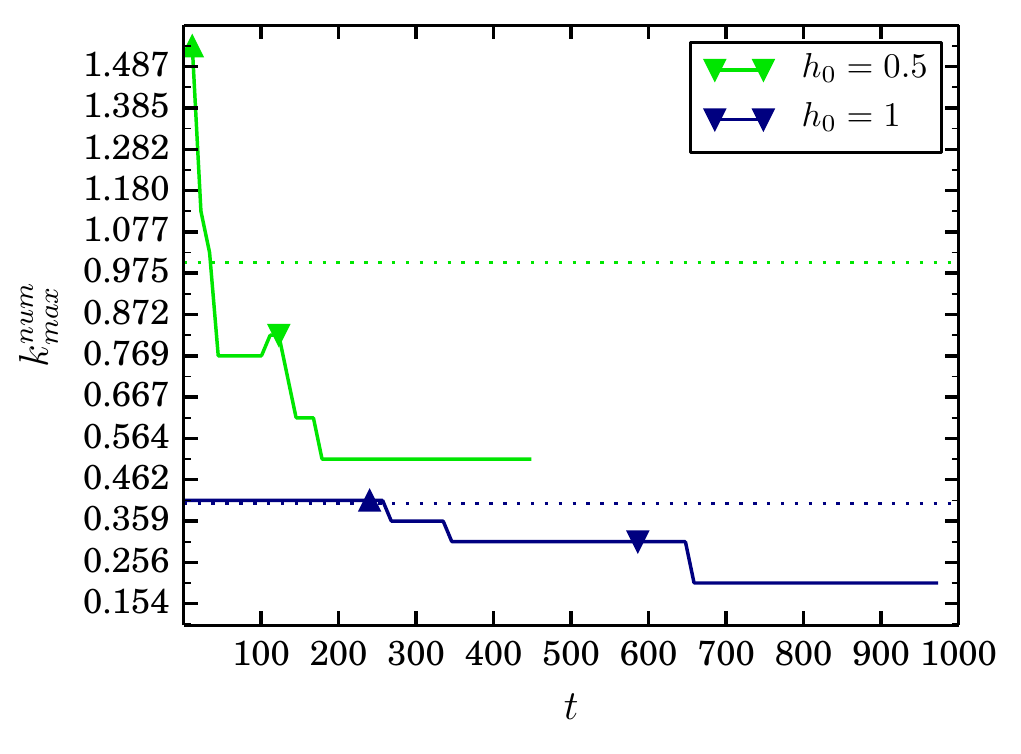}
	\caption{Comparison of $k_{\text{max}}^{\text{num}}$ as a function of
	time for 3D films with initial thickness $h_0=1$ (blue) and $h_0=0.5$
	(green).
	The blue curve is identical to that seen in Fig.~\ref{fig:kmax_3D}.
	The blue (green) triangles show the times at which the first holes
	appear, and the inverted ones show the times at which droplets begin to form.
	The blue (green) dashed lines show the corresponding values of 
	 $k_{\text{max}}$.   $\mbox{Oh}=0.487$, and $\theta_{\text{eq}}=0.439\pi$ for
	both simulation sets.}
	  \label{fig:3dkmaxcomparison}
\end{figure}

Fig.~\ref{fig:3dkmaxcomparison} shows the time evolution of $k_{max}^{num}$ for 3D films with
 $h_0=1$ and $h_0=0.5$. 
Similarly to the case when $h_0=1$ plotted in Fig.~\ref{fig:kmax_3D},
$k_{max}^{num}$ shifts to smaller wavenumbers after two
distinct events: first, when holes begin to form in the film,
and second when drops begin to form (marked by triangles
and inverted triangles, respectively).
However, two significant differences are observed for the thinner film.
First, breakup takes place almost immediately, so there is no
interval of time where $k_{max}^{num}$ approximates $k_{max}$.
Second, for larger times, the difference between $k_{max}$ and $k_{max}^{num}$
is greater for $h_0=0.5$ than for $h_0=1$.

To summarize, the DFT of the nonlinear film breakup shows that the dominant
length scales deviate from $\lambda_{\text{max}}$ predicted by the LSA.  The degree of
deviation depends on $\theta_{\text{eq}}$, and whether the film is 2D or 3D;
additionally, the time it takes for the film to rupture depends strongly on
$\mbox{Oh}$.  For 2D simulations with small $\theta_{\text{eq}}$, the DFT has a
peak at or close to $k_{\text{max}}$ for the entirety of the film evolution.
For larger $\theta_{\text{eq}}$, the evolution of the DFT shows two distinct
phases: prior to breakup, when the peak in the DFT is at $k_{\text{max}}$, and
after breakup, when the peak shifts to smaller wavenumbers.  For 3D
simulations, the evolution of the DFT shows three distinct phases, each
associated with a shift towards smaller wavenumbers: prior to breakup, after
holes begin to form, and after drops begin to form.  A decrease in $\mbox{Oh}$
is primarily associated with a dramatic increase in the time it takes for the
film to rupture. 

\clearpage
\section{Conclusions}\label{sec:conclusions}

In this paper, we have demonstrated a computationally efficient method
for including fluid-solid interactions into direct numerical simulations.
This method is found to perform as well as the body force formulation~\cite{MahadyvdW15}, 
while requiring a significantly smaller
computational effort.   The two methods were compared by considering 
contact angles of equilibrium drops in 2D, where both methods perform similarly in
terms of convergence with mesh refinement, as well as in terms of the behavior
when the equilibrium film thickness, $h^\ast$, is reduced.
The computational complexity required is however dramatically reduced for the
presently considered method.

With the established improvement in computational performance, we are now able to  
study the instability of films due to fluid-solid interaction
using direct numerical simulations to gain a quantitative understanding
of the evolution of the instability, film rupture, and the post-rupture dewetting
process.  We compare the results of direct simulations with the linear stability
analysis (LSA) of the long-wave formulation (L-W),
and find that when contact angle, $\theta_{\text{eq}}$, is larger, there is a significant
difference with the predictions of the LSA.
We also demonstrate that a reduction in the film thickness does not 
reduce the differences between the LSA and direct simulations, as the typical 
in-plane lengthscale decreases even faster than the out-of-plane one (film thickness).  

Finally, we study breakup of a film in both 2D and 3D. 
We consider the evolution of the emerging length scales by computing the discrete Fourier
transform (DFT) of the fluid profile.  2D films are characterized by a two stage evolution,
and 3D films by a three stage one; for both cases,
the initial phase exhibits a DFT with a peak, $k_{\text{max}}^{\text{num}}$, near $k_{\text{max}}$
from the LSA, and each successive stage is associated with
a decrease in $k_{\text{max}}^{\text{num}}$, and correspondingly, an increase
in the emerging length scales.
When a flat film is perturbed by small perturbations, the simulations illustrate
the dynamics of dewetting, the formation of `dry spots', and the subsequent coarsening 
due to inertial effects. When studying the effect of the contact angle for 2D films, we find that the dewetting
morphology depends on the contact angle; particularly, the agreement between the simulation 
results and the predicted fastest growing unstable mode from the LSA deteriorates for large contact 
angles after the formation of the first expanding holes.
When studying the breakup of films in the presence of significant inertial effects, so that characteristic  
Ohnesorge number, $\mbox{Oh}$, is small, we find that the long term preference for a specific wavenumber 
is weaker when compared with the larger $\mbox{Oh}$, resulting
in an increased degree of disorder in the distribution of emerging patterns.

The speed and simplicity of the implementation of the method presented in this paper opens up
the possibility of studying a variety of problems involving thin films and other geometries that have
not been studied extensively so far by direct numerical simulations. Our numerical method allows, 
for the first time, for the simulation of arbitrary contact angles (including those greater than $\pi/2$),
and film breakup, including inertial effects.
While we have studied the influence of varying contact angle and $\mbox{Oh}$ in 2D films,
we leave the exhaustive parameter study of 3D films for a future work.\\

{\bf Acknowledgments} 
This work was partially supported by the NSF Grants DMS-1320037 and CBET-1235710.


\end{document}